\documentclass[aps, prd, amsmath, amssymb, amsfonts, floats, %
floatfix, superscriptaddress, nofootinbib, twocolumn, showpacs]%
{revtex4}

\allowdisplaybreaks[2]  

\usepackage{graphicx}
\usepackage[usenames, dvipsnames]{color}
\usepackage[colorlinks, pdfborder={0 0 0}, plainpages=false]{hyperref}
\usepackage{breakurl}
\usepackage{amsmath, amssymb, amsfonts}
\usepackage{xspace} 
\usepackage{dcolumn}
\usepackage{bm}
\usepackage{multirow} 

\def\laq{\raise 0.4ex\hbox{$<$}\kern -0.8em\lower 0.62ex\hbox{$\sim$}}
\def\gaq{\raise 0.4ex\hbox{$>$}\kern -0.7em\lower 0.62ex\hbox{$\sim$}}

\definecolor{CiteColor}{rgb}{0, 0.5, 0} %
\hypersetup{citecolor=CiteColor} %
\definecolor{RefColor}{rgb}{0.55, 0, 0} %
\hypersetup{linkcolor=RefColor} %

\usepackage{ulem}
\newcommand{\Caltech}{\affiliation{Theoretical Astrophysics 350-17,
    California Institute of Technology, Pasadena, CA 91125, USA}}

\newcommand{\Cornell}{\affiliation{Center for Radiophysics and Space
    Research, Cornell University, Ithaca, NY, 14853, USA}}
\newcommand{\Fullerton}{\affiliation{Gravitational Wave Physics and Astronomy Center,
    California State University Fullerton, Fullerton, CA 92831, USA}}

\newcommand{\e}[1]{\ensuremath{\times 10^{#1}}}
\newcommand{\sub}[1]{_{\rm #1}}
\newcommand{\super}[1]{^{\rm #1}}

\newcommand{\roughly}{\mathchar"5218\relax} 
\newcommand{\aref}[1]{Appendix~\ref{#1}}

\newcommand{\run}[2]{\ensuremath{S^{#1#1}_{#2}}}
\newcommand{\Ch}{Christodoulou\xspace}
\newcommand{\MCh}{\ensuremath{M_{\rm Ch}}\xspace}
\newcommand{\Erad}{\ensuremath{E_{\rm rad}}\xspace}
\newcommand{\fourth}{fourth}
\newcommand{\second}{second}
\newcommand{\third}{third}
\newcommand{\fifth}{fifth}

\begin{document}

\title{Final spin and radiated energy in numerical simulations of binary black holes with equal masses and equal, aligned or anti-aligned spins}
 
\author{Daniel A. Hemberger} \Cornell
\author{Geoffrey Lovelace} \Fullerton\Caltech
\author{Thomas J. Loredo} \Cornell
\author{Lawrence E. Kidder} \Cornell
\author{Mark A. Scheel} \Caltech
\author{B\'ela Szil\'agyi} \Caltech
\author{Nicholas W. Taylor} \Caltech
\author{Saul A. Teukolsky} \Cornell

\begin{abstract} 
The behavior of 
merging black holes (including the emitted gravitational waves and the 
 properties of the remnant) can currently
be computed only by
numerical simulations.
This paper introduces ten numerical relativity simulations of
binary black holes with equal masses and equal spins aligned or anti-aligned
with the orbital angular momentum. The initial spin magnitudes have
$|\chi_i| \lesssim 0.95$ and are more concentrated in the aligned
direction because of the greater astrophysical interest
of this case.
We combine these data
with five previously reported simulations of the same configuration,
but with different spin
magnitudes,
including the
highest spin simulated to date, 
$\chi_i\approx0.97$. This data set
is sufficiently accurate to
enable us to offer improved analytic 
fitting formulae for the final spin and for the energy
radiated by gravitational waves as a function of initial spin.
The improved fitting formulae can help to improve our understanding 
of the properties of binary black hole merger remnants and  
can be used to enhance future
approximate waveforms for gravitational wave searches,
such as Effective-One-Body waveforms.
\end{abstract}

\date{\today}


\maketitle

\section{Introduction}
\label{sec:intro}

Binary black holes are an important source for gravitational-wave
detectors such as
the Laser Interferometer Gravitational-Wave
Observatory (LIGO), GEO, and 
Virgo~\cite{Abbott:2007, Grote:2008zz, Acernese:2008}.
Searches
for gravitational-wave signals
have been able to constrain the event rate
for binary black hole mergers, but a direct
detection of gravitational
waves has not yet been made~\cite{Abadie:2011kd, 2012arXiv1209.6533T}. 
These searches require predictions (``templates'') of the 
expected gravitational waves;
so far, only non-spinning templates have been
included~\cite{2012arXiv1209.6533T}.
However,
there is evidence that spin is 
relevant in astrophysical black
holes, from both theoretical
predictions~\cite{Thorne:1974,GammieEtAl:2004,KesdenEtAl:2010} and
observational data~\cite{McClintockEtAl:2006,WangEtAl:2006,Brenneman:2011wz}.

Therefore, 
LIGO and other gravitational-wave
detectors need to include spin
as a parameter in their template waveforms; otherwise, the search space
(and thus the detection rate)
is reduced
because of an insensitivity to spinning sources
\cite{AjithEtAl:2009,VanDenBroeck:2009gd}.
Accurate simulations of spinning binary black hole mergers are
also needed to infer the properties (e.g. masses and spins) of
binaries from the detected waveforms
(``parameter estimation")~\cite{Aasi:2013kqa}.

For both detection and parameter estimation, 
numerical simulations are too computationally expensive to 
generate waveforms for
the entire parameter space of binary black hole mergers.
Instead, 
numerical simulations are used to calibrate and validate 
the approximate, analytic models 
that are actually used to generate template waveforms.
For instance, the Effective-One-Body (EOB) model, 
calibrated using numerical simulations that include merging black holes 
with spins aligned or anti-aligned 
with the orbital angular momentum~\cite{Taracchini:2012},
is used by the LIGO Collaboration to estimate how
sensitive their search is to waveforms from spinning 
systems~\cite{2012arXiv1209.6533T}.
However, Ref.~\cite{Taracchini:2012} has shown that the EOB model poorly
predicts configurations
with large aligned spins, and that more
numerical relativity simulations are needed in this region 
of spin parameter space to improve the calibration of the model.

Binary black holes whose spins are aligned (or anti-aligned) with the
orbital angular momentum involve far fewer parameters than generic
binaries with arbitrary spin directions, but nevertheless
they can be used to construct templates capable of
detecting a sizeable fraction of precessing binaries~\cite{AjithEtAl:2009}.
Furthermore, aligned-spin
systems are astrophysically motivated by studies including
observations of the micro-quasar XTE J1550-564~\cite{Steiner:2011vr},
models of gas-rich galaxy mergers~\cite{Bogdanovic:2007hp}, and
population synthesis models~\cite{Fragos:2010tm}.

In this paper, we introduce ten new simulations of binary black holes
with equal masses and equal spins aligned or anti-aligned with
the orbital angular momentum.
We use the notation
$S_{|\chi|}^{\pm\pm}$ to refer to specific cases, where the subscript is
approximately the dimensionless spin magnitude at $t=0$, and the superscripts
indicate whether each black hole has
the aligned ($+$) or anti-aligned ($-$) spin orientation.
The new simulations are \run+{0.95}, \run+{0.9}, \run+{0.85}, \run+{0.8},
\run+{0.6}, \run+{0.2}, \run-{0.2}, \run-{0.6}, \run-{0.8}, and \run-{0.9}.

To more fully cover the aligned-spin space, we include data
previously reported for \run+{0.97}~\cite{Lovelace:2011nu},
\run-{0.95}~\cite{Lovelace:2010ne}, \run-{0.0}~\cite{Garcia:2012dc},
\run-{0.44}~\cite{Chu2009}, and
\run+{0.44}~\cite{Thesis:Chu} in our analysis.
The \run+{0.95} case joins the two simulations from
Refs.~\cite{Lovelace:2010ne,Lovelace:2011nu} as the only simulations to
date of merging black holes with spin magnitudes above
$\chi\approx0.93$ (the ``Bowen-York 
limit")~\cite{cook90,DainEtAl:2002,HannamEtAl:2009}.
We use this combined dataset to improve on prior
phenomenological fitting formulae for the
final spin of the merger remnant~\cite{Rezzolla:2007xa,Barausse2009,Tichy2008}
and the radiated energy from inspiral through
ringdown~\cite{Tichy2008,Reisswig:2009vc,Barausse:2012qz}.
These improved formulae can be used to reduce EOB waveform phase
errors in the ringdown
(see Eq.~(19) and surrounding text in Ref.~\cite{Taracchini:2012})
and therefore provide more accurate
templates for gravitational-wave searches~\cite{Barausse:2012qz}.

The remainder of this paper is organized as follows.
In Sec.~\ref{sec:Methods}, we discuss the numerical methods that we employ in
our simulations.
In Sec.~\ref{sec:Simulations}, we report 
on the
values and convergence of the constraint
violations, masses, and spins.
In Sec.~\ref{sec:Results}, we use the horizon data to improve the
phenomenological fitting formulae for final spin and radiated energy
as a function of initial spin.
Section~\ref{sec:Conclusions} contains our conclusions,
and~\aref{app:model} details our method for constructing the 
fitting formulae.

\section{Simulation methods}
\label{sec:Methods}


All simulations used in this paper were generated with the
Spectral Einstein Code (SpEC)~\cite{SpECwebsite}. In this
section, we describe the methods for the ten
new simulations.
For detailed methods of the previously
reported SpEC simulations,
see Refs.~\cite{Lovelace:2011nu,Lovelace:2010ne,Garcia:2012dc,Chu2009,Thesis:Chu}
and references therein.
Throughout this paper, we use units where $G=c=1$, and we
report lengths and
times in units of $M$, the total Christodoulou mass in the initial data.

To produce initial data, we solve
the extended conformal thin-sandwich equations 
with quasi-equilibrium boundary 
conditions~\cite{York1999,Cook2002,Cook2004,
  Caudill-etal:2006,Gourgoulhon2001,Grandclement2002}.
We adopt free data 
based on a weighted superposition of two Kerr-Schild black holes,
which enables us to construct initial data 
containing black holes with nearly extremal spins~\cite{Lovelace2008,Lovelace2009}.
The constraint equations are solved using
a spectral elliptic solver~\cite{Pfeiffer2003}, and the 
free parameters are iterated
until the target masses and spins are achieved to within some tolerance.

We evolve the initial data
on a ``cut-spheres" domain~\cite{Buchman:2012dw}
using spectral adaptive mesh
refinement, which will be detailed in a forthcoming paper.
On a timescale of $50M$, we change smoothly from the
initial data gauge to damped harmonic
gauge~\cite{Lindblom2009c,Choptuik:2009ww,Szilagyi:2009qz}, which helps prevent
coordinate singularities. We use a fifth-order Dormand-Prince dense adaptive
time stepper.

To reduce eccentricity, we first evolve each system for 2.5 orbits beyond 
the time when the
spurious ``junk'' radiation is sufficiently far from the black holes so as to have a
negligible effect on the black hole trajectories.
Then we fit
the time derivative of the orbital frequency
$\dot{\Omega}$ to find improved initial angular and radial velocities
($\Omega_0$ and $\dot{d}_0/d_0$)~\cite{Mroue2010}.
We iterate this procedure until
an eccentricity below $10^{-3}$ is achieved.

\begin{figure}
\begin{center}
\includegraphics[bb=0 0 4800 2820, width=3.4in]{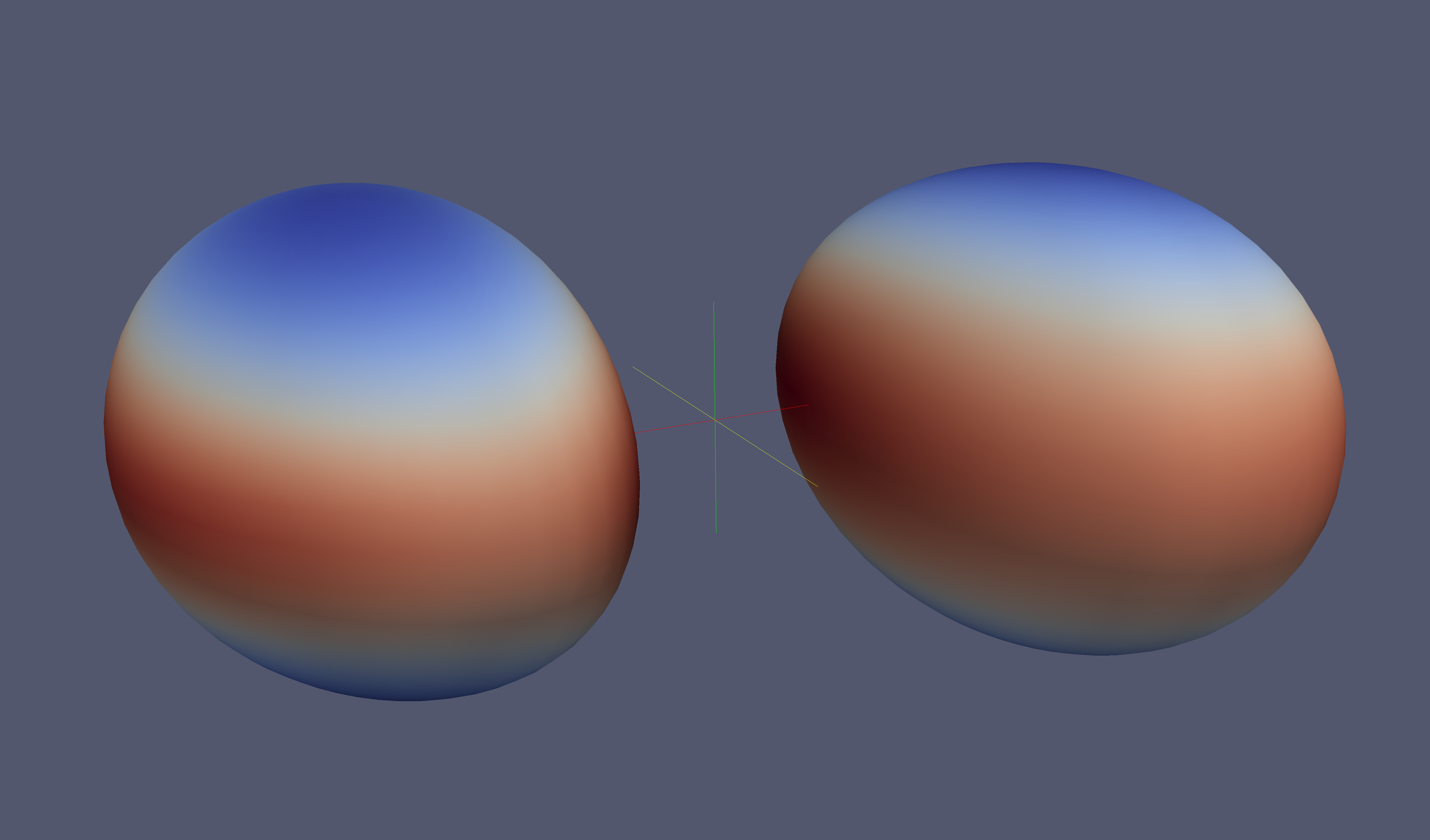}
\end{center}
\caption{Effect of spin on horizon geometry.
This image shows the
intrinsic
Ricci scalar on the apparent horizons in 
simulation \run+{0.85}. The proper separation of the horizons
along the line connecting their centers is about $1.7M$.
Both spin effects (gradients as a function of polar angle)
and tidal bulges (dark red regions near the intersection with
the line connecting the horizon centers) can be seen.
\label{fig:ricci}}
\end{figure}

We use the dual-frames technique to do spectral excision of
the singularities~\cite{Hemberger:2012jz}.
As described in other papers reporting high spin simulations using
SpEC~\cite{Lovelace:2011nu,Lovelace:2010ne}, the most important
aspect of this excision technique is careful control of the
excision boundary. This must accomplish three tasks. First,
it must distort the shape of the boundary so that it matches the
shape of the apparent horizon. Second, it must regulate the
fractional separation between the excision surface and the apparent
horizon --- if the separation is too small, then the horizon falls
out of the computational domain, but if the separation is too large,
then the excision surface falls far inside the horizon, where
large gradients are computationally expensive to resolve.
Third, it must keep all characteristic speeds on the excision
surface positive; i.e., the excision surface must be a pure outflow 
boundary. This is because we do not impose boundary conditions on
the excision surface.
Instead, we monitor the characteristic speeds; if 
they ever become negative,
then our evolution system
becomes ill-posed,
and we terminate the simulation.
These three tasks are challenging for high spin systems in part because
of the additional distortion of the horizons
(see Fig.~\ref{fig:ricci}), and
they are especially challenging for large aligned spins because 
such systems
spend more time in the dynamic regime before merger.

%


Using the fast-flow method described in Ref.~\cite{Gundlach1998},
we find the apparent horizons as an expansion
in spherical harmonics, truncated at a given maximum order $\ell$.
As the system evolves, we adaptively change $\ell$ to
satisfy accuracy criteria for the resolution of the horizon.
After a common horizon is found during merger,
the evolution continues on
a new domain with a single excision surface
that subsumes the two individual excision regions~\cite{Szilagyi:2009qz}.

We measure the quasi-local spin, $S$, on each horizon using the approximate
Killing vector
method described in Ref.~\cite{Lovelace2008}. The dimensionless spin is then
\begin{equation}
\chi = \frac{S}{M_{\rm Ch}^2},
\end{equation}
where \MCh is the \Ch mass,
\begin{equation}
M_{\rm Ch}^2 = M_{\rm irr}^2 + \frac{S^2}{4M_{\rm irr}^2},
\end{equation}
and $M_{\rm irr}$ is the irreducible mass, which is
a function of the horizon area,
\begin{equation}
M_{\rm irr} = \sqrt{\frac{A_{\rm AH}}{16\pi}}.
\end{equation}
With these definitions, $\chi=1$ represents an extremal black
hole~\cite{BoothFairhurst:2008}.

We choose an integer $k$ to characterize the resolution of each simulation.
We call $k$ the resolution level (or ``Lev"). It sets the
resolution by defining the target maximum truncation error for the
adaptive mesh refinement and adaptive horizon finding as
\begin{equation}
\epsilon_{\rm max} = 10^{-4}e^{-k}.
\end{equation}
Around the excision boundary, where
the most resolution
is required, we reduce the target maximum truncation error for
adaptive mesh refinement by a factor of $10^{2}$.

\section{Simulations}
\label{sec:Simulations}

There are ten new simulations presented in this paper: \run+{0.95},
\run+{0.9}, \run+{0.85}, \run+{0.8}, \run+{0.6}, \run+{0.2}, \run-{0.2},
\run-{0.6}, \run-{0.8}, and \run-{0.9}. Initial data were generated
with a target \Ch mass
for each hole $M_0=0.5$, target spin for each hole $\chi_0$, 
and target ADM linear
momentum $p_0^i = 0$. We fix the initial separation at $d_0 = 15.366 M$,
and then we iterate as summarized
in Sec.~\ref{sec:Methods} to obtain the initial
radial velocity $\dot{d}_0/d_0$ and angular velocity $\Omega_0$.
The targets are met to within an absolute error of 
$\mathcal{O}(10^{-8})$, and the
resulting initial data parameters are reported in Table~\ref{table:ID}.
We construct our initial data with a target total \Ch mass of $M=1$ 
so that
our evolution code units are essentially interchangeable
with units of $M$.

\begin{table}
\begin{center}
\begin{tabular}{c || c | c | c}
$\chi_0$ & $M\dot{d}_0/d_0 \e{4}$ & $M\Omega_0$ & $N_{\rm orbits}$ \\
\hline
0.95   &  7.26420673 & 0.01395360 & 25.4 \\
0.9    &  5.48222492 & 0.01419573 & 24.9 \\
0.85   &  4.33347923 & 0.01437107 & 24.7 \\
0.8    &  3.54332917 & 0.01450430 & 24.2 \\
0.6    &  1.65215665 & 0.01487274 & 22.8 \\
0.2    &  0.09507527 & 0.01525060 & 19.9 \\
-0.2   &  -0.69081937 & 0.01538827 & 17.2 \\
-0.6   &  -1.95883097 & 0.01527384 & 14.6 \\
-0.8   &  -3.60252091 & 0.01501397 & 13.3 \\
-0.9   &  -5.42657163 & 0.01474328 & 12.8 \\
\end{tabular}
\end{center}
\caption{Initial data parameters (radial velocity $\dot{d_0}$ and
angular velocity $\Omega_0$) at separation $d_0=15.366M$
for the ten new simulations with target
spin, $\chi_0$. Also included is the approximate number of orbits until 
merger. Here $M$ is the sum of the Christodoulou masses 
at time $t=0$.
}
\label{table:ID}
\end{table}

\begin{figure}
\begin{center}
\includegraphics[bb=0 0 576 432, width=3.5in]{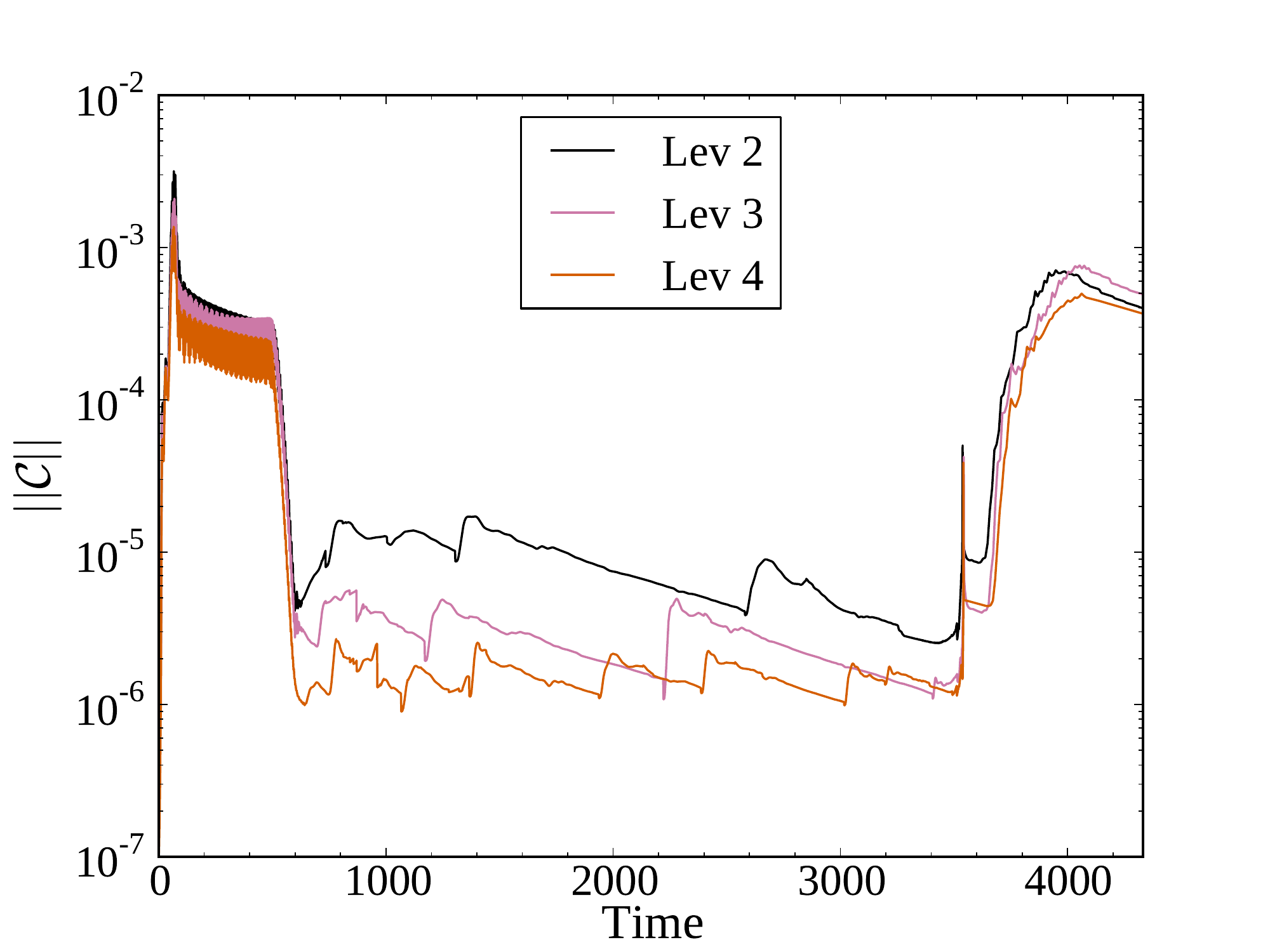}
\end{center}
\caption{
Normalized constraint violations for \run-{0.9}.
For each resolution level, $k$, we plot $||\mathcal{C}||$,
the volume-averaged $L^2$-norm of
the generalized harmonic constraint energy divided by the 
volume-averaged
$L^2$-norm of the dynamical field gradients. This 
measure is defined
in Eq.~(71) of Ref.~\cite{lindblom2006}.
As the resolution level increases, the constraints decrease. 
Jumps in the
constraints are attributed to changes in
the domain structure,
and the spike around $t \sim 3500 M$ corresponds to the merger.
}
\label{fig:constraints}
\end{figure}

At least three different resolutions were evolved
for each case to check convergence.
Figure \ref{fig:constraints}
shows convergence
of the (normalized) volume-averaged $L^2$-norm of the 
generalized harmonic constraint energy~\cite{lindblom2006}
for a representative
case\footnote{
Note that we observe poor convergence of the constraints in the late
ringdown, which indicates that the simulations may need to be re-run
to provide accurate waveforms.
However, this issue occurs in the wave zone well after the
black hole has settled down to its final mass and spin,
and so it does not affect the results of this paper.
}.
Each time the domain structure is changed to alleviate grid compression,
the constraints jump because of interpolation errors,
but then slowly decay back to their baseline levels.

Additional simulations are used in our analysis of masses and spins
in Sec.~\ref{sec:Results}:
\run+{0.97}~\cite{Lovelace:2011nu},
\run+{0.44}~\cite{Thesis:Chu},
\run-{0.0}~\cite{Garcia:2012dc},
\run-{0.44}~\cite{Chu2009}, and
\run-{0.95}~\cite{Lovelace:2010ne}.
Although these have been
previously reported, we include 
them
below for completeness.
It should be noted that \run+{0.44} and \run-{0.44} are
older simulations
and therefore used different initial data and evolution
machinery than described in Sec.~\ref{sec:Methods}.
Simulations \run-{0.95}, \run-{0.0}, and \run+{0.97} used 
the initial data methods of Sec.~\ref{sec:Methods}, but 
earlier implementations of the evolution methods.

\subsection{Mass and spin}
\label{sec:MassAndSpin}

\begin{figure}
\begin{center}
\includegraphics[bb=0 0 576 432, width=3.5in]{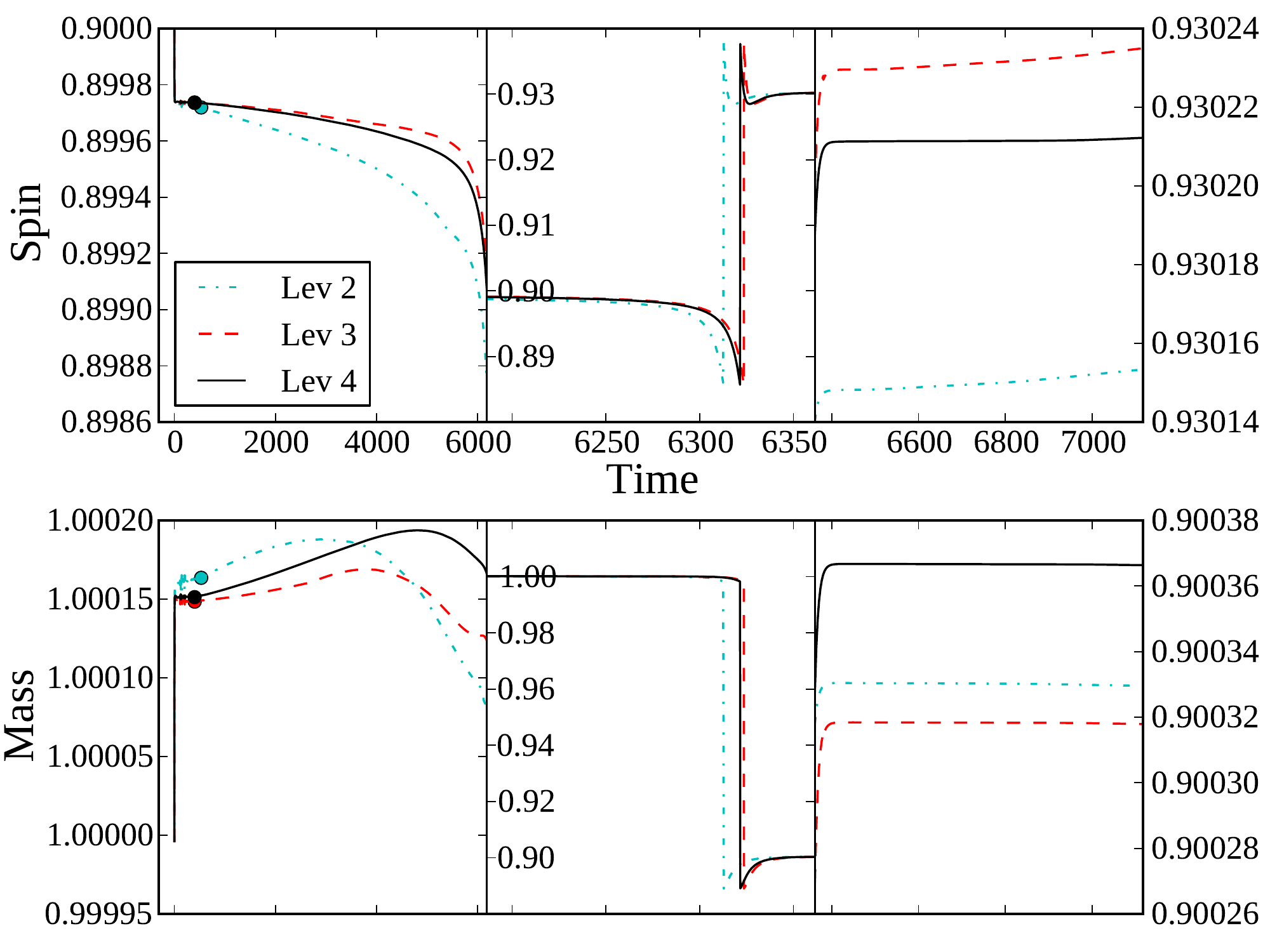}
\end{center}
\caption{
Plots of the apparent horizon quantities as a function of
time for a representative case, \run+{0.9}.
The top panels display the dimensionless spin and the bottom panels
display the \Ch mass. From left to right, the panels display the
inspiral, merger, and ringdown. We normalize the $y$-scales
separately so that the differences between each resolution can
be clearly seen. The discontinuity
in the middle panel indicates where we begin to
measure the mass and spin on the common horizon.
The dots in the early inspiral identify our choice of $t_i$
for each resolution level.
}
\label{fig:AhVsTime}
\end{figure}

We define the initial spin, $\chi_i$, to be the spin after the system has
relaxed from the initial data and the junk radiation at
the apparent horizon has become negligible. The spin before this time
is not physically relevant to the rest of the evolution.
There are subtle issues to consider when choosing the time
to measure $\chi_i$.
If we choose too early a
time, then junk radiation effects will still be present. If we are overly
cautious and choose too late a time, then the system will have emitted
enough gravitational radiation to significantly change the spin.

We use a histogram method to determine $\chi_i$.
Let $\{\chi_I\}$ be the set of spin measurements during the
inspiral. The range of $\{\chi_I\}$
is split uniformly into $N$ bins, where $N$ is the size of $\{\chi_I\}$, and
then each element of $\{\chi_I\}$ is put into the appropriate bin.\footnote{
If the time interval between spin measurements is not equally spaced, we
weight each measurement by the average of the two adjacent time intervals.}
We choose $\chi_i = \chi(t_i)$,
where $t_i$ is the latest time when the
spin is in the bin containing the most measurements.
In the initial relaxation, the spin is oscillating, and during the inspiral,
the spin changes more rapidly as the holes approach each other. 
Under these conditions, this
method selects the spin just after the junk radiation, when the spin is nearly
constant for a long interval.
Figure~\ref{fig:AhVsTime} shows the evolution of the mass and spin as a
function of time for a representative case,
and identifies our choice of $t_i$ by the dots in the early inspiral.

\begin{table}
\begin{center}
\begin{tabular}{c || c | c}
Case & $|\chi_i|$ & $|\chi_f|$ \\
\hline
\run-{0.95}  & 0.949053(-30)  & 0.37567(-18)  \\
\run-{0.9}  & 0.899569(-11)  & 0.392748(-12)  \\
\run-{0.8}  & 0.7997602(59)  & 0.4268932(30)  \\
\run-{0.6}  & 0.59993163(71)  & 0.4942327(-31)  \\
\run-{0.44}  & 0.437568970(-10)  & 0.547851(20)  \\
\run-{0.2}  & 0.1999802(-40)  & 0.6242202(-61)  \\
\run-{0.0}  & 64(-29)\e{-8}  & 0.686445(-52)  \\
\run+{0.2}  & 0.200035(-19)  & 0.7464314(-96)  \\
\run+{0.44}  & 0.4365505(95)  & 0.8140(10)  \\
\run+{0.6}  & 0.5999635(14)  & 0.857808(15)  \\
\run+{0.8}  & 0.7998737(-44)  & 0.907526(14)  \\
\run+{0.85}  & 0.849826(15)  & 0.919088(30)  \\
\run+{0.9}  & 0.8997371(-15)  & 0.930212(23)  \\
\run+{0.95}  & 0.9495863(-25)  & 0.940852(29)  \\
\run+{0.97}  & 0.969504(13)  & 0.944964(11)  \\
\end{tabular}
\end{center}
\caption{Dimensionless spin measurements. For each case,
we provide the initial spin magnitude
of each hole and the final spin magnitude of the remnant
at the highest resolution.
Note that the number in parentheses is not an error estimate,
but provides the value at the next highest resolution
when added to the last two significant digits.
}
\label{table:Spin}
\end{table}

We compute $t_i$ from $\chi$ because
the behavior of the mass is not as simple during the inspiral.
The histogram method applied to the mass will pick out the local maximum late
in the inspiral that is present in most of our cases
(see Fig.~\ref{fig:AhVsTime}). We define
the initial mass
to be $M_i = M(t_i)$, the
sum of the
\Ch masses at time $t_i$.

The final spin and \Ch mass, $\chi_f$ and $M_f$, are
measured at the last observation time, when the merger remnant is in 
quasi-equilibrium and approximates a Kerr black hole.
We report the initial and final spins in
Table~\ref{table:Spin}, and the initial and final \Ch masses
in Table~\ref{table:Mass}.

\begin{table}
\begin{center}
\begin{tabular}{c || c | c || c}
Case & $M_i$ & $M_f$ & $\Erad (\%)$ \\
\hline
\run-{0.95} & 0.999856(68) & 0.968134(33) & 3.1727(33) \\
\run-{0.9} & 1.00016197(73) & 0.967909(-27) & 3.2248(28) \\
\run-{0.8} & 1.0000859(-11) & 0.9665941(-16) & 3.348894(50) \\
\run-{0.6} & 1.00002292(-78) & 0.963769(-14) & 3.6253(13) \\
\run-{0.44} & 2.2470608(-22) & 2.159561(-49) & 3.8940(21) \\
\run-{0.2} & 0.999956(26) & 0.9564388(84) & 4.3519(17) \\
\run-{0.0} & 0.9999971(-43) & 0.9516182(-74) & 4.83791(33) \\
\run+{0.2} & 0.999961(22) & 0.945471(16) & 5.44923(46) \\
\run+{0.44} & 2.2451548(28) & 2.10099(-44) & 6.421(20) \\
\run+{0.6} & 1.00001907(-96) & 0.926868(-19) & 7.3149(18) \\
\run+{0.8} & 1.0000765(-22) & 0.911275(-28) & 8.8794(26) \\
\run+{0.85} & 1.000108(-12) & 0.906168(-73) & 9.3931(62) \\
\run+{0.9} & 1.0001513(-29) & 0.900366(-48) & 9.9770(46) \\
\run+{0.95} & 1.00021743(77) & 0.893703(-65) & 10.6492(66) \\
\run+{0.97} & 1.0002384(-94) & 0.890691(-22) & 10.9521(14) \\
\end{tabular}
\end{center}
\caption{
\Ch mass measurements. For each case,
we provide the total initial mass of the black holes,
the final mass of the remnant, and the radiated energy
computed from Eq.~\eqref{eq:Erad}
at the highest resolution.
Note that the number in parentheses is not an error estimate,
but provides the value at the next highest resolution
when added to the last two significant digits.
}
\label{table:Mass}
\end{table}

From the initial and final \Ch masses,
we can infer the fraction of
the black hole energy that is radiated
in gravitational waves
during the evolution:
\begin{equation}
\Erad = 1 - \frac{M_f}{M_i}.
\label{eq:Erad}
\end{equation}

\begin{figure}
\begin{center}
\includegraphics[bb=0 0 576 432, width=3.5in]{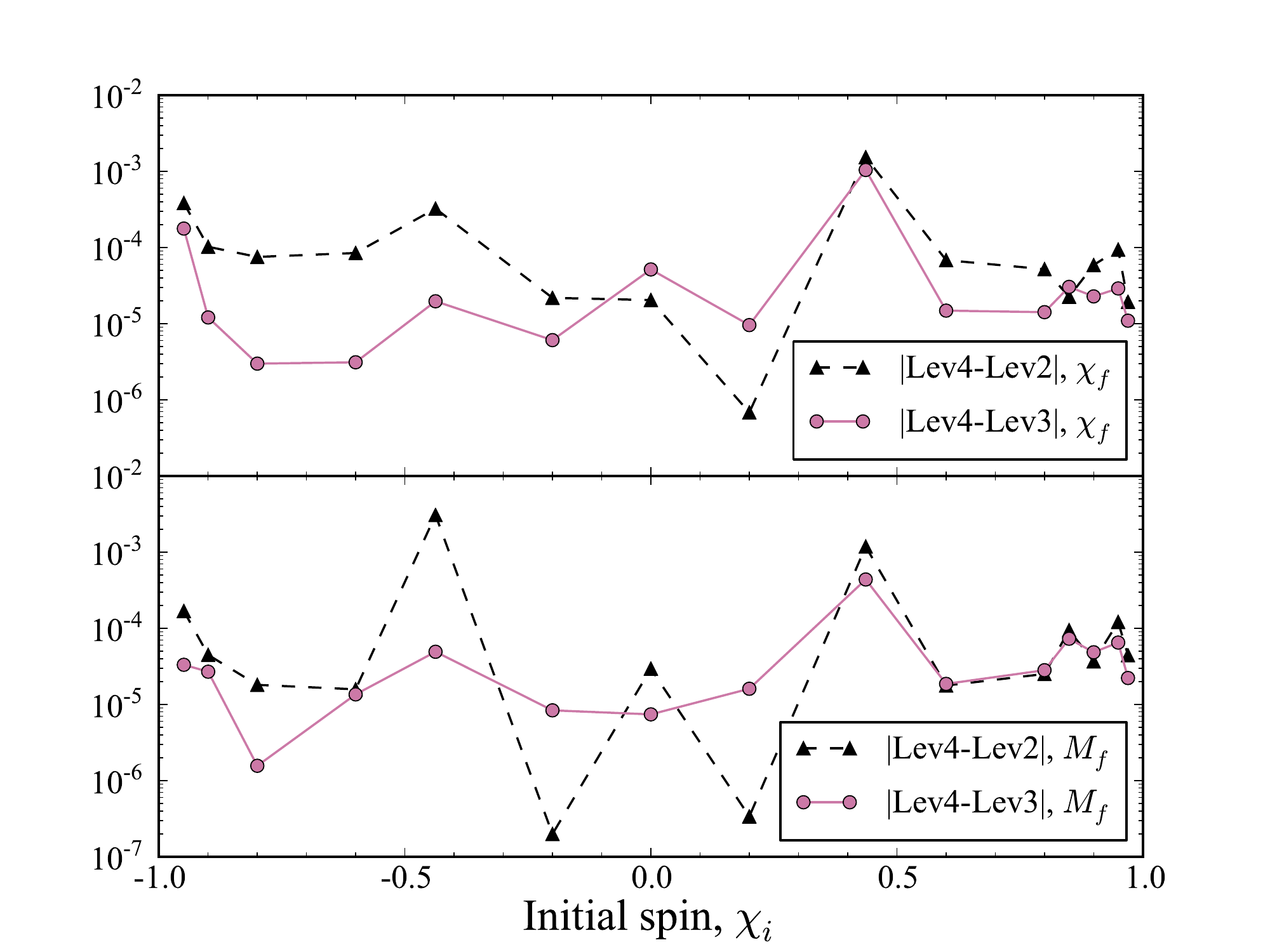}
\end{center}
\caption{
Differences in the final masses and spins between resolution levels.
For each case, we compare $\chi_f$ and $M_f$ of the
highest resolution to the two lower resolutions.
Note that, except for the older \run\pm{0.44} simulations,
all differences are $\lesssim 10^{-4}$.
Differences in the initial masses and spins behave similarly.
}
\label{fig:MassSpinConvg}
\end{figure}

We expect mass and spin measurements at higher resolutions to
be more accurate.
However, as illustrated by the comparisons
in Fig.~\ref{fig:MassSpinConvg},
these quantities are not strictly convergent 
in a number of cases.
For this reason, we include measurements from all resolutions
in our analysis in Sec.~\ref{sec:Results} and
weight the uncertainty in the error assigned to a particular
measurement by a function of resolution level $k$.

As described in Sec.~\ref{sec:Methods}, the most stringent 
resolution
requirements occur in the vicinity of
the apparent horizons, but the accuracy
may be dominated by short, under-resolved segments of the evolution.
The initial masses and spins appear to be randomly perturbed by
the junk radiation as the initial data relax.
The final masses and spins,
which are already limited by the accuracy of the initial masses and spins,
appear to also be affected by the details of the coalescence,
where we see a spike in constraint violations
(Fig.~\ref{fig:constraints}).
Apart from these under-resolved segments, we do see convergence in the
time derivatives of the masses and spins, but the absolute values remain
offset from one another.

\begin{figure}
\begin{center}
\includegraphics[bb=0 0 576 432, width=3.5in]{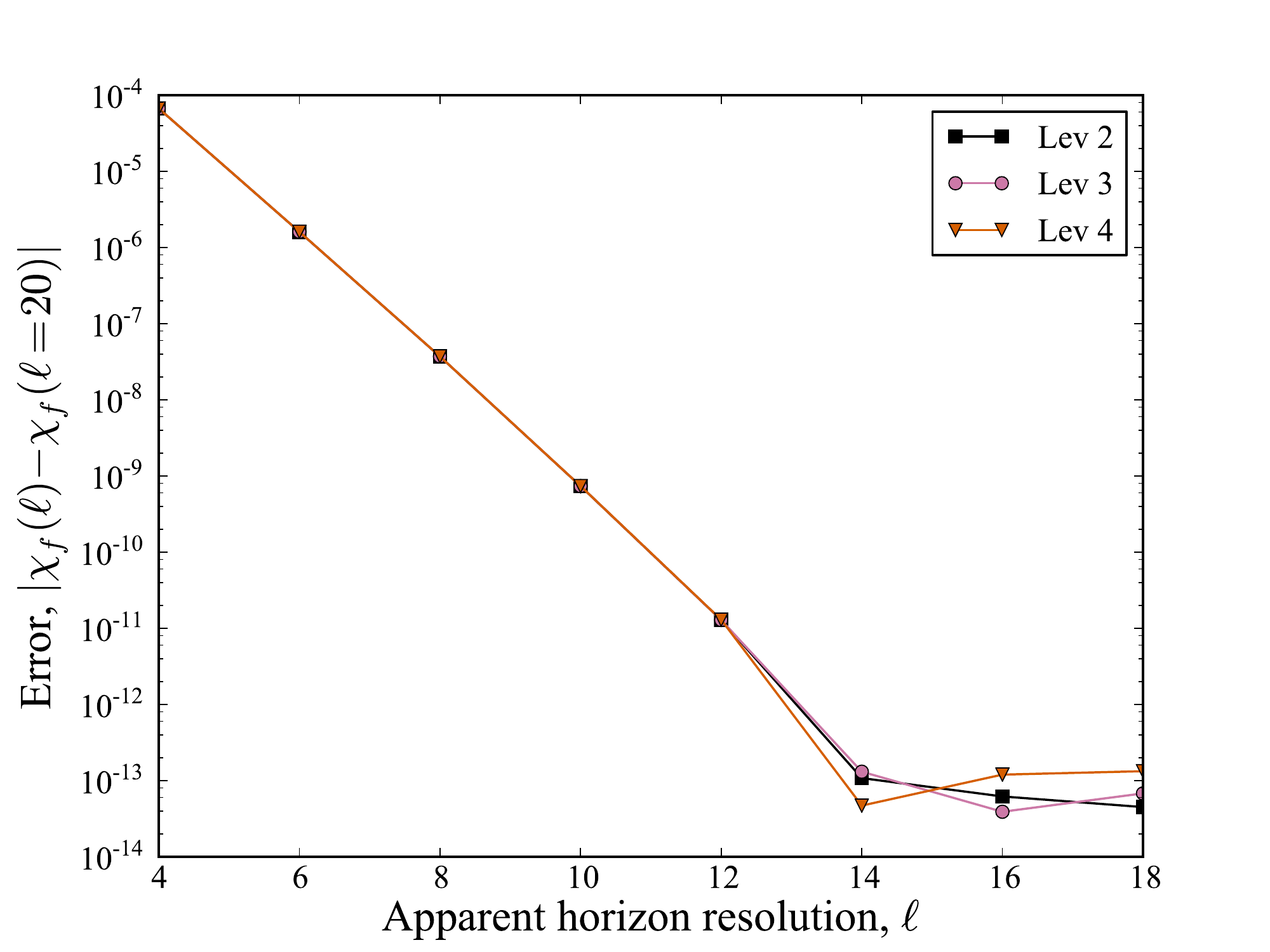}
\end{center}
\caption{
Convergence of the final spin for \run-{0.9} as a function
of the $\ell$ of the horizon finder. We plot the difference
between $\chi_f(\ell)$ and $\chi_f(\ell=20)$ for each resolution.
The adaptive horizon finder for this case chose $\ell=8$
at the final time of the simulation.
}
\label{fig:LConvg}
\end{figure}

We have investigated other potential sources of uncertainty, but found
them to lie below the resolution level uncertainty.
For example, one source of uncertainty in the masses and spins is the resolution of the
surface of the horizon.
In Fig.~\ref{fig:LConvg}, we show a representative plot of error in final
spin as a function of $\ell$ of the horizon finder. 
Let $\Delta\chi_k$ be the resolution level error between
the two highest
resolutions, and let $\Delta\chi_{\ell}$ be the resolution error
between the $\ell$ chosen by the adaptive horizon finder and
the $\ell$ for which the horizon is fully resolved.
At the final time of the simulation, we find that, in all cases,
$\Delta\chi_k > \Delta\chi_{\ell}$ by several orders of magnitude.

A source of uncertainty in the radiated energy
is the energy that would have
been radiated by the binary as it proceeds from infinite
separation to the separation $d_0$ at which we start the simulation.
As discussed in Ref.~\cite{Lovelace:2011nu}, Alvi's formula~\cite{Alvi:2001mx} estimates
that the energy radiated from $d=\infty$ to $d=d_0$ is one part in $10^6$.
Since this is smaller than our resolution level uncertainty, it is
safe to ignore this difference and we can think of \Erad as
the total radiated energy from infinite separation through ringdown.

\section{Results}
\label{sec:Results}

Much effort has been put into constructing phenomenological formulae for the
final spin~\cite{Campanelli2006d,Rezzolla:2007xa,Barausse2009,Tichy2008}
and radiated energy~\cite{Campanelli2006d,Tichy2008,Reisswig:2009vc,Barausse:2012qz}
as a function of initial spin.
Because the SpEC code has the capability to generate and evolve initial data of
black holes with spins above the ``Bowen-York limit"
of $\chi\approx0.93$~\cite{HannamEtAl:2009}, we are able
to provide new data points to test and improve these formulae.

We use a Bayesian nonlinear measurement error model (described in \aref{app:model})
to fit and compare new parametric formulae.  This approach (1)~accounts for
uncertainties in both the initial spin data and the
output data (i.e., final spin or radiated energy); (2)~accounts for
the expected improvement in accuracy of results as the resolution level
increases; and (3)~includes a simple systematic error component quantifying
misfit between a chosen formula and the curve the data are
converging toward.\footnote{
To keep the calculations analytically tractable, the systematic error
component accounts only for the typical magnitude of misfit (essentially,
the root-mean-square of the residuals), and does not
account for correlations or patterns in
the residuals.
}
The framework lets us predict an output
as a function of initial spin, with prediction uncertainties that account
for the uncertainties in the parameters of the chosen formula (including
correlated uncertainties) and the typical scale of the systematic error.

The new fitting formulae that we provide here are only applicable
to equal mass binary black hole configurations with
equal spins aligned or anti-aligned
with the orbital angular momentum. More general formulae
exist (see e.g. Refs.~\cite{Lousto2009,Gergely:2012wm,Barausse2009,Tichy2008}),
but they
are less accurate at high spins because of the
scarcity of simulations with both unequal masses and high spins in random
orientations.

\subsection{Final spin}
\label{subsec:Finalspin}

\begin{figure}
\begin{center}
\includegraphics[width=3.5in]{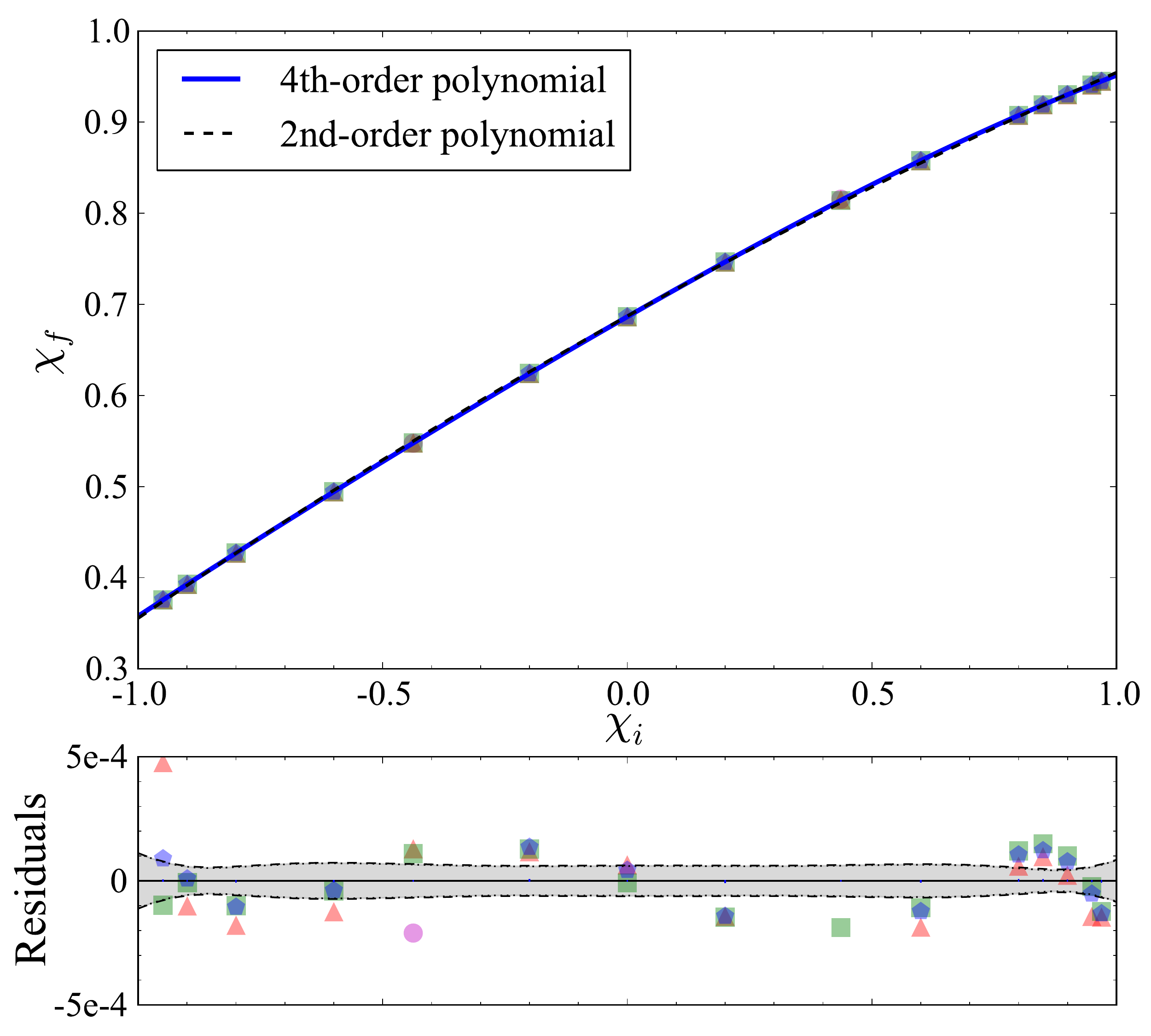}
\end{center}
\caption{
In the top panel, we plot our preferred fitting formula (solid line),
the \fourth-order polynomial in Eq.~\eqref{eq:FinalSpinFit},
and a comparison with a \second-order polynomial (dashed line)
for $\chi_f$ as a function of $\chi_i$.
Our data points are plotted as polygons, where
more sides indicates higher resolution level.
In the bottom panel, we plot the residuals of the \fourth-order polynomial.
We indicate our fit parameter (dotted line) and total prediction (dashed line)
uncertainties (defined in \aref{app:model}),
which in this case are nearly identical.
Note that the
residuals for the two lower resolution
runs for \run+{0.44} are too large to fit in this panel.
}
\label{fig:FinalSpin}
\end{figure}

Using the data from Table~\ref{table:Spin}, we construct a new
fitting formula for the final spin
as a function of initial spin.
We fit to a \fourth-order polynomial,
\begin{equation}
\tilde{\chi}_f = a_0 + a_1\chi_i + a_2\chi_i^2 + a_3\chi_i^3 + a_4\chi_i^4.
\label{eq:FinalSpinFit}
\end{equation}
The best fit to our data has the parameters $a_n$ and associated covariance $\Sigma_a$:
\begin{align}
\label{eq:FinalSpinCoefs}
a &= \left(\begin{array}{r} a_0 \\ a_1 \\ a_2 \\ a_3 \\ a_4 \end{array}\right)
=
\left(
\begin{array}{r}
0.686402(60) \\
0.30660(14)  \\
-0.02684(33) \\
-0.00980(19) \\
-0.00499(35)
\end{array}
\right) \\
\Sigma_a & = \left(
\begin{array}{rrrrr}
   3.6  &   0.31 &  -14    &  -0.45 &   11  \\
   0.31 &  21    &   -4.8  & -26    &    6.0 \\
 -14    &  -4.8  &  110    &   7.1  & -110   \\
  -0.45 & -26    &    7.1  &  36    &   -9.5 \\
  11    &   6.0  & -110    &  -9.5  &  120
\end{array}
\right)\e{-9}
\end{align}
The uncertainty in $a_n$, given
in parentheses in Eq.~(\ref{eq:FinalSpinCoefs}), is estimated by $\sqrt{\Sigma^{nn}_a}$.
However, the parameter estimates are highly correlated;
therefore, the full covariance matrix is used
in the computation of the fit uncertainty $\sigma_f$
in Eq.~\eqref{sigma-f}.
In Fig.~\ref{fig:FinalSpin}, we show the fit and residuals
using Eq.~\eqref{eq:FinalSpinFit} with the parameters
from Eq.~\eqref{eq:FinalSpinCoefs}.

We fit to a \fourth-order polynomial because the high accuracy of our dataset
enables us to identify significant \third- and \fourth-order 
trends in the residuals of a
fit to a \second-order polynomial, which is the fitting function used
in Refs.~\cite{Rezzolla:2007xa,Barausse2009,Tichy2008}.
The difference between the logarithm of the marginalized likelihood function (LML)
for the best-fit \fourth-order
and \second-order polynomials is $\roughly 40$, indicating that the \fourth-order
polynomial provides a significantly better fit.
If the two additional degrees of freedom were fitting
noise, rather than some underlying structure in the data, then we
would only expect a change in maximum LML of $\mathcal{O}(1)$.\footnote{
The leading-order term in the
maximum LML is proportional to a chi-squared-like quantity,
so for nested models, such as the \second-
and \fourth-order polynomials,
the change in the maximum LML
should roughly mimic the asymptotic statistics of likelihood
ratio tests, as given by Wilks's theorem~\cite{Wilks1938}.  Two models are said
to be nested if the simpler one is a special case of the more
complicated one.}

The estimated systematic error magnitude $\hat{\sigma}_\Delta$
(defined in \aref{app:model})
for the \fourth-order polynomial
formula is negligibly small, suggesting that the significant behavior
is captured as well as could be expected.  
However, the
residuals, especially at large aligned spins, display trends suggesting
that there is additional structure not captured by the
\fourth-order polynomial (such trends are ignored by our simple
systematic error model).  This encouraged us to explore a \fifth-order
polynomial formula, but it did not reduce the residuals
enough to justify the additional degree of freedom.  This does
not rule out the possibility that a different
formula could capture the behavior even more 
accurately.

\begin{figure}
\begin{center}
\includegraphics[bb=0 0 576 432, width=3.5in]{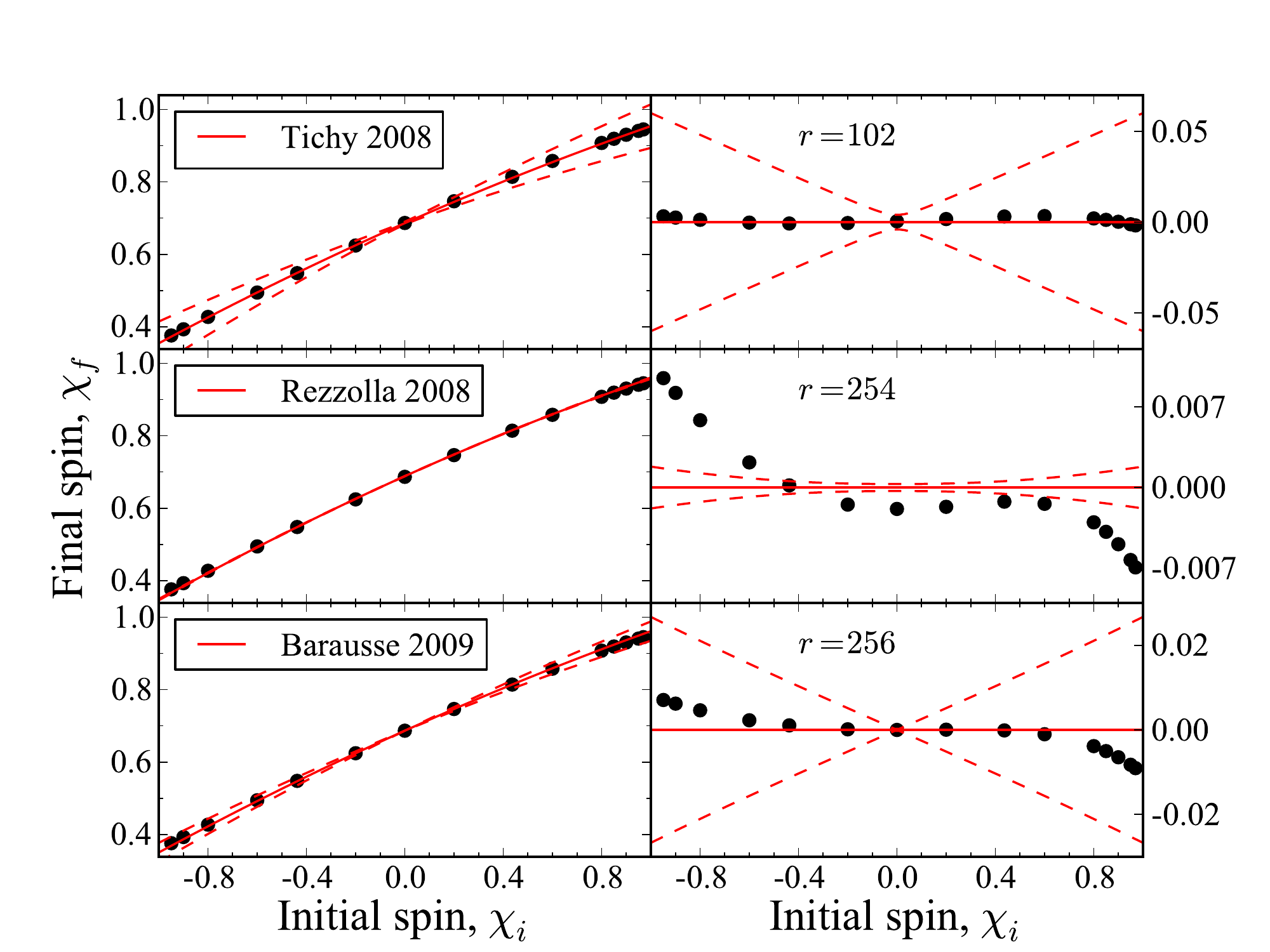}
\end{center}
\caption{
Final spin as a function of initial spin.
In the left panels we plot our data (circles)
along with the fitting formula (red line)
with error estimates (dashed) from several other 
studies. The top panel
is from Ref.~\cite{Tichy2008}, the middle is 
from Ref.~\cite{Rezzolla:2007xa},
and the bottom is from Ref.~\cite{Barausse2009}.
In the right panels we plot the difference between our data and
the corresponding fitting formula on the
left. The value $r$ quantifies
 the size of the systematic
error compared to the \fourth-order polynomial.
}
\label{fig:FinalSpin_compare}
\end{figure}

We compare our data to existing fitting formulae for the final spin
in Fig.~\ref{fig:FinalSpin_compare}.
The $\chi_f$ data corroborate the existing fitting formulae, but
indicate deviations at large spins (especially in the aligned
direction).
This is an expected consequence of the scarcity of high spin
numerical relativity data heretofore.
In the figure, we provide a quantity for each fit, $r$, that
measures how much larger its systematic error is than
that of our \fourth-order polynomial.
This is essentially a ratio of the
root-mean-square residuals
(more precisely, $r$ is the ratio of the
$\hat{\sigma}_\Delta$ values).
The previously reported formulae have roughly 100
to 250 times as much systematic error as our \fourth-order
polynomial fit. 
While the formula in Tichy 2008~\cite{Tichy2008}
performs best, we note that it has a large uncertainty.

\subsection{Radiated energy}

\begin{figure}
\begin{center}
\includegraphics[width=3.5in]{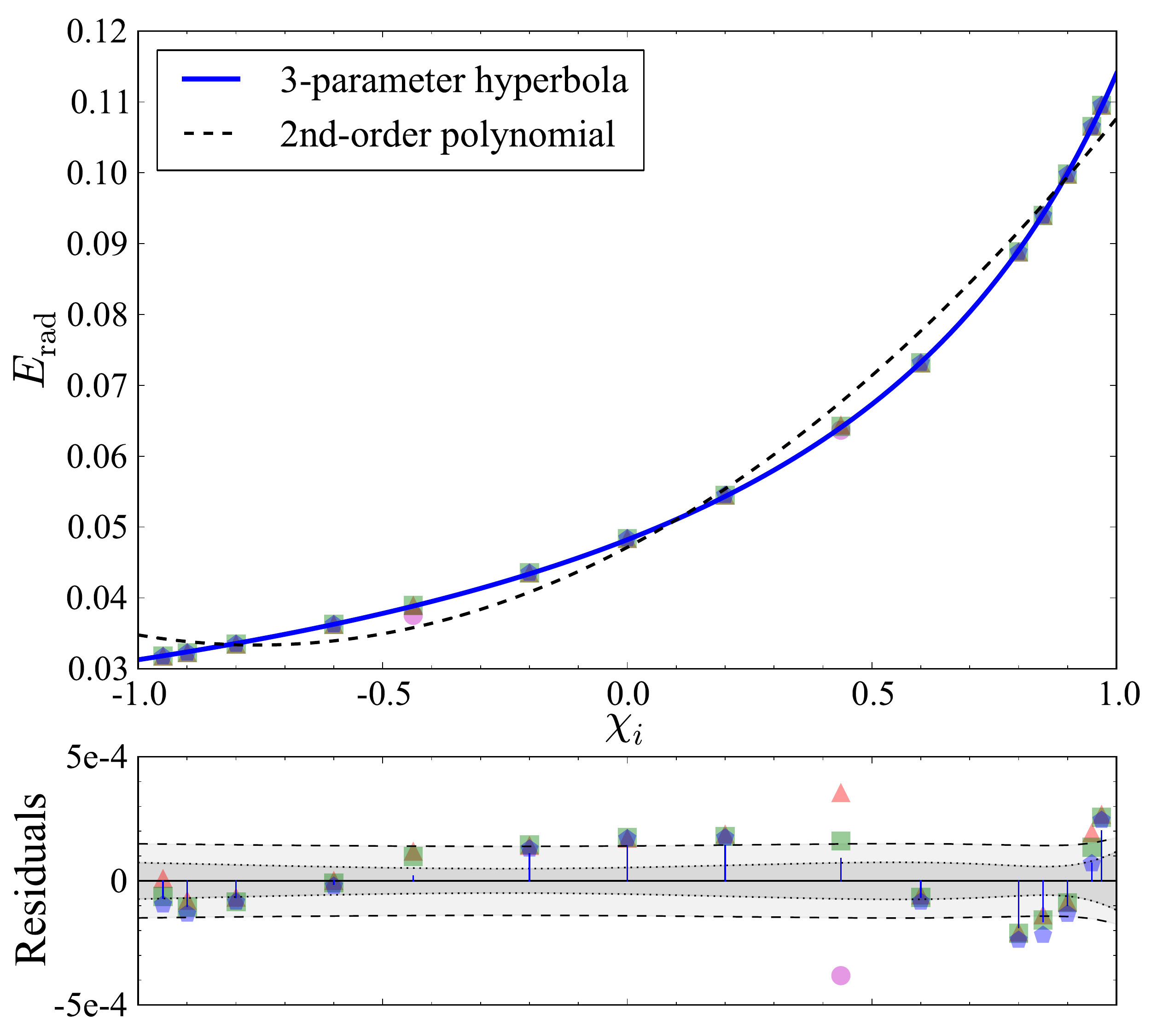}
\end{center}
\caption{
In the top panel, we plot our preferred fitting formula (solid line),
the hyperbolic function in Eq.~\eqref{eq:EradFit},
and a comparison with a \second-order polynomial (dashed line) for \Erad
as a function of $\chi_i$.
Our data points are plotted as polygons, where more sides indicates
higher resolution level.
In the bottom panel, we plot the residuals of the hyperbolic function.
We indicate our fit parameter (dotted line)
and total prediction (dashed line) uncertainties (defined in \aref{app:model}).
}
\label{fig:Erad}
\end{figure}

Following the procedure in Sec~\ref{subsec:Finalspin}, we
use the data from Table~\ref{table:Mass} to construct a new
fitting formula for the radiated energy fraction, \Erad, as
a function of initial spin.
We fit to a hyperbolic function,
\begin{equation}
\tilde{E}_{\rm rad} = b_0 + \frac{b_1}{b_2 + \chi_i}.
\label{eq:EradFit}
\end{equation}
The best fit to our data has the parameters $b_n$ and associated covariance $\Sigma_b$:
\begin{align}
\label{eq:EradCoefs}
b &= \left(\begin{array}{r} b_0 \\ b_1 \\ b_2 \end{array}\right)
=
\left(
\begin{array}{r}
0.00258(29) \\
-0.07730(79)\\
-1.6939(59)
\end{array}
\right) \\
\Sigma_b &= \left(
\begin{array}{rrr}
  0.83 &  2.2 &  16 \\
  2.2 &  6.2 &  46 \\
  16 &  46 & 350
\end{array}
\right)\e{-7}
\end{align}
The uncertainty in $b_n$,
given in parentheses in Eq.~(\ref{eq:EradCoefs}), is estimated by $\sqrt{\Sigma^{nn}_b}$.
In Fig.~\ref{fig:Erad}, we show the fit and residuals
using Eq.~\eqref{eq:EradFit} with the parameters
from Eq.~\eqref{eq:EradCoefs}.

We use a hyperbolic fitting function instead of a
\second-order polynomial
(as in Refs.~\cite{Tichy2008,Reisswig:2009vc})
or a constrained \second-order polynomial, e.g.
$\tilde{E}_{\rm rad} = c_0 + c_1\chi_i + (c_1/4)\chi_i^2$
(as in Ref.~\cite{Barausse:2012qz}).
Parabolic fits show visible offsets in various regions of the initial
spin space, which can be seen
in plots of the residuals in~\cite{Barausse:2012qz} and in the comparison
plot in the top panel of Fig.~\ref{fig:Erad}.
The difference between the
maximum LML for the 3-parameter hyperbola and the \second-order
polynomial is $\roughly 36$, 
indicating that the hyperbola is a dramatically
better fit to the data.

\begin{figure}
\begin{center}
\includegraphics[bb=0 0 576 432, width=3.5in]{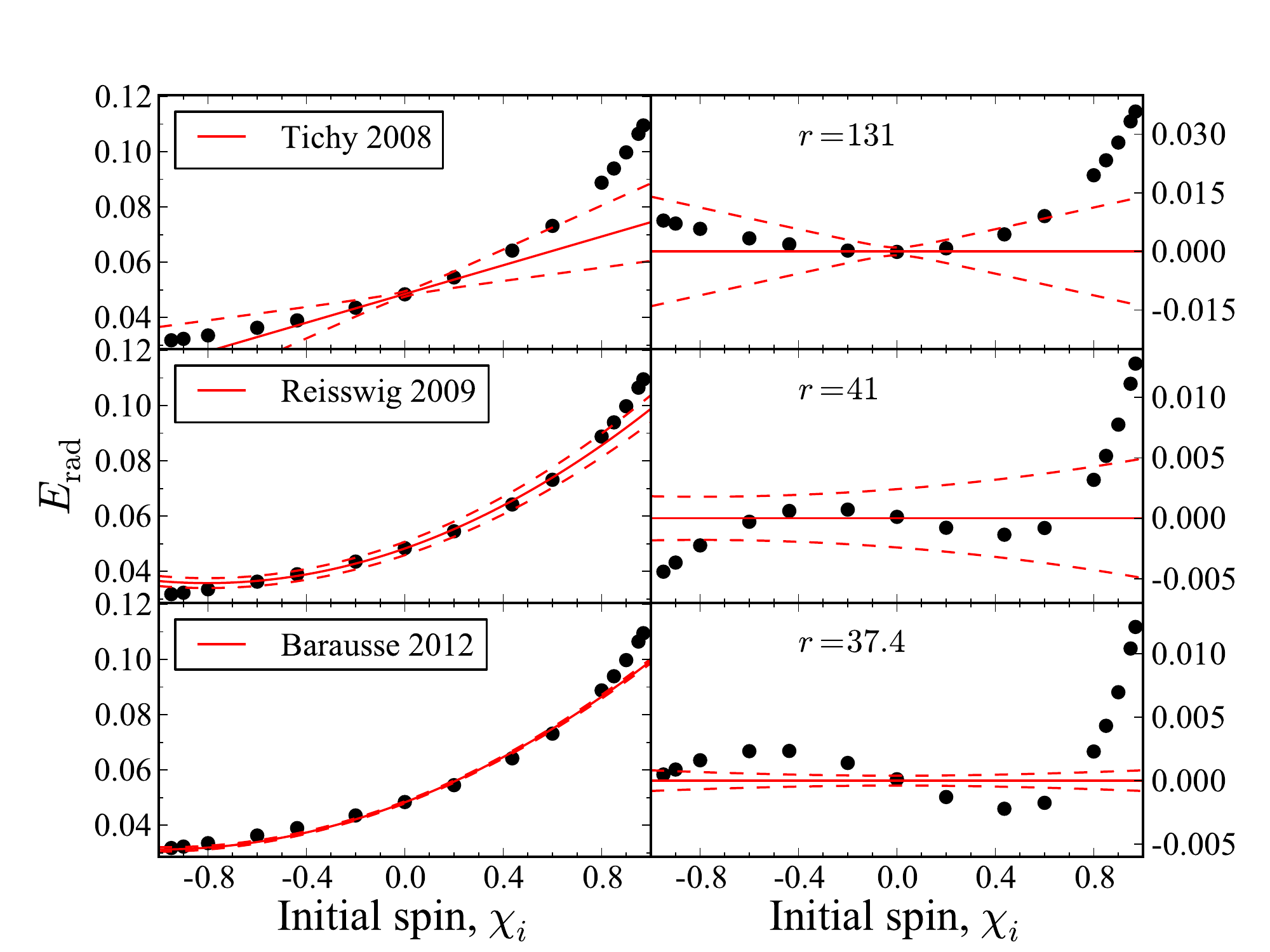}
\end{center}
\caption{
\Erad as a function of initial spin.
In the left panels we plot our data (circles) along with the
fitting formula (red line)
with error estimates (dashed) from several other studies.
The top panel
is from Ref.~\cite{Tichy2008}, 
the middle is from Ref.~\cite{Reisswig:2009vc},
and the bottom is from Ref.~\cite{Barausse:2012qz}.
In the right panels we plot the difference between our data and
the corresponding fitting formula on the left.
The value $r$ quantifies the size of the systematic
error compared to the 3-parameter hyperbola.
}
\label{fig:Erad_compare}
\end{figure}

In Fig.~\ref{fig:Erad_compare},
we compare our data to existing fitting formulae for \Erad.
All previous formulae suffer from the same systematic deficiencies
as the best \second-order polynomial fit to our
data shown in Fig.~\ref{fig:Erad}.
The ratio of the systematic error magnitude in these formulae 
to its magnitude in
our 3-parameter hyperbolic fit, $r$, is shown in the
figure and ranges from roughly 40 to 130.
Note that it is not meaningful to compare these $r$ values to those
shown in Fig.~\ref{fig:FinalSpin_compare}, because
 we have not added
any additional degrees of freedom in our \Erad model compared
to a \second-order polynomial (unlike in our final spin model,
which adds two degrees of freedom).

\subsection{Extremality}

An important aspect of these fitting formulae is their ability to predict
remnant properties for nearly extremal initial spins.
How much of the initial mass can be radiated as gravitational waves, 
and how fast can the remnant hole spin?
Our prediction for the radiated energy and final spin for an extremal
initial spin configuration, $\chi_i=1$, is
\begin{align}
\tilde{E}_{\rm rad}(1) &= 0.11397(18), \\
\tilde{\chi}_f(1) &= 0.951383(85),
\end{align}
where the uncertainty (in parentheses)
is $\sigma\sub{tot}$, the total prediction uncertainty defined
in Eq.~\eqref{sigma-tot}, evaluated at $\chi_i=1$.
The highest radiated energy predicted by any of the formulae
we compare against in this paper
is $\Erad(1) = 0.0995(8)$.

Previous estimates of \Erad underestimated
the mass loss for large, aligned initial spins.  The most extreme data point
in this paper, \run+{0.97}, was identified as a potential
outlier~\cite{Barausse:2012qz}
and \Erad was expected to be $\lesssim 10\%$ for an extremal, aligned
configuration inspiraling from infinity.
Additional data presented here, most notably \run+{0.95} and \run+{0.9},
suggest that \run+{0.97} is not an outlier. Furthermore, these cases indicate
that even a $\chi_i=0.9$ inspiral is capable of radiating $\gtrsim 10\%$ of
its initial mass.

Simulations with $\chi_i>0.93$
are an important factor in our fitting formulae. 
This high-spin regime is not accessible with the most popular 
initial data methods 
for binary black hole evolutions, 
which assume conformal flatness (cf. Ref.~\cite{Lovelace2008} and 
references therein). 
To assess the impact of the high spin simulations,
we compare our best fits to fits of a subset of the data,
omitting cases \run-{0.95}, \run+{0.95}, and \run+{0.97}.
We identify several key results.

For the $\chi_f$ formula, we find that
$\Delta\tilde{\chi}_f(1) \gtrsim 2.5\sigma\sub{tot}\super{sub}(\chi_f,1)$.
That is, the prediction of the final spin with the full dataset differs
from the prediction with the subset by more than 2.5 times the total
prediction uncertainty in the fit to the subset.
The parameter uncertainty in the full dataset is smaller by a factor
$\sigma_f/\sigma_f\super{sub} \approx 0.6$, which is much smaller than 
would be
expected from adding 3 data points
randomly distributed in initial spin
(the expected improvement based on the root-$N$ rate would be
$\sqrt{12/15}\approx 0.9$).
In a random sampling context, one would typically have to
more than double the size of the dataset to get such a
reduction in parameter uncertainties.  Of course, we have
not chosen the subset randomly. 
Note that because
the systematic error magnitude is negligible,
$\hat{\sigma}_\Delta\approx 0$, 
the total prediction 
uncertainty of Eq.~\eqref{sigma-tot} has the same
behavior as the parameter uncertainty.

For the \Erad formula, we find that
$\Delta\tilde{E}\sub{rad}(1) \gtrsim 3.5\sigma\sub{tot}\super{sub}(\Erad,1)$;
the lower spin subset poorly predicts the extremal \Erad.
Parameter uncertainties decrease only slightly faster than the expected
root-$N$ rate for adding 3 randomly placed data points,
$\sigma_f/\sigma_f\super{sub} \approx 0.85$. However, the total prediction
uncertainty at $\chi_i=1$ increases,
$\sigma\sub{tot}/\sigma\sub{tot}\super{sub} \approx 1.15$,
because the additional high spin data deviate most from the fitting formula
that is based on lower spin data. That is, \Erad for $\chi_i > 0.93$ is
unanticipated by the fit to the lower spin subset, causing the systematic
error magnitude $\hat{\sigma}_\Delta$ to increase.
While this highlights the importance of the high spin data in assessing
the predictive power of the fitting formula for near-extremal initial spins,
it also suggests that we are unlikely to capture the behavior of \Erad
much better with our simple fitting formula.
Furthermore, neither manual nor algorithmic searches~\cite{Schmidt03042009}
have identified any better formulae,
which leads us to believe that for
the best predictive results
at high spins, a non-parametric approach may be preferred.
Such an approach could be implemented using, for example,
a correlated Gaussian process~\cite{MR2514435},
which would provide a way to predict final masses and spins
without the use of a parametric fitting formula.

The analysis comparing the subset to the full dataset does not change in any
appreciable way if we use a near-extremal spin, e.g. $\chi_i=0.97$, instead
of the most extreme case, $\chi_i=1$.


%
%

\section{Conclusions}
\label{sec:Conclusions}


In this paper, we present and analyze
a family of numerical relativity simulations performed using 
SpEC
in order to construct improved fitting formulae for the final spin
and radiated energy as a function of initial spin. We consider
a physically
motivated, one-dimensional subset of the binary black hole parameter
space, in which the black holes have
equal masses and equal spins aligned or anti-aligned with
the orbital angular momentum.
The improvement in these fitting formulae is most dramatic
in the regime where the initial spin is above the
``Bowen-York limit," since for the first time
data from simulations above this limit have been included in the fits.

For the final spin, we improve on the \second-order polynomial fitting formula
by using a \fourth-order polynomial to capture the statistically significant
cubic and quartic features.
For the radiated energy,
we find that a 3-parameter hyperbolic fitting formula is 
greatly
preferred to a \second-order polynomial.
The qualitatively different behavior at large, aligned spins in the
new fit to \Erad implies
that there is somewhat more power in gravitational waves from nearly extremal
sources than previously thought, perhaps because of
higher-order effects that become relevant at very high spins.

We have shown that performing
more nearly extremal simulations is
the most effective way to reduce
the uncertainty in the fitting formula parameters. 
However, we have also 
observed that
the systematic uncertainty in \Erad 
may prohibit a simple fitting formula
from providing any further significant improvement to the prediction 
uncertainty
of \Erad for high, aligned spins.


Analytic models, such as the aligned-spin EOB model, are needed to
generate templates for gravitational-wave detectors (e.g. LIGO),
because of the prohibitive expense of generating sufficient numerical
relativity data to adequately cover the parameter space.
The fitting formulae we define in this paper can be used to
better calibrate these models, and therefore improve future
template waveforms.

\newcommand{\model}{\mathcal{M}}
\newcommand{\modelset}{\{\model_\n\}}
\newcommand{\params}{\theta}
\newcommand{\paramvec}{\vec{\theta}}
\newcommand{\pspace}{{\cal P}}
\newcommand{\like}{\mathcal{L}}
\newcommand{\mlike}{\mathcal{L}_m}  
\newcommand{\plike}{\mathcal{L}_p}  
\newcommand{\olike}{\ell}  
\newcommand{\Dobs}{D_{\rm obs}}
\newcommand{\dvec}{\bm{d}}
\newcommand{\expect}{\mathbb{E}}
\newcommand{\Normal}{\mathop{\mathtt{Norm}}\nolimits}  
\newcommand{\normal}{\mathcal{N}}
\newcommand{\eps}{\epsilon}
\newcommand{\ispin}{\chi_i}
\newcommand{\fspin}{\chi_f}
\newcommand{\dmm}{E_{\rm rad}}
\newcommand{\xtrue}{\xi}
\newcommand{\ytrue}{\eta}
\newcommand{\fparams}{\theta}
\newcommand{\hypers}{\psi}
\newcommand{\xerr}{\epsilon}
\newcommand{\yerr}{\delta}
\newcommand{\syserr}{\Delta}
\newcommand{\n}{n}
\renewcommand{\k}{k}

\appendix
\section{Method for constructing our fitting formulae}
\label{app:model}

Our goal is to find a convenient but reasonably accurate function that
predicts the final black hole spin, $\fspin$, or the fractional radiated energy,
$\dmm$, as a function of the initial spin, $\ispin$.  We will
specify one or more simple parametric candidate functions,
find the best parameter values, quantify uncertainties in the
parameters and predictions, and compare rival candidate functions.  To
treat both the $\fspin$ and $\dmm$ problems in generic notation, we let
$\xtrue$ denote the {\em predictor} (i.e., $\ispin$), and $\ytrue$ denote the
{\em response} we seek to predict (i.e., $\fspin$ or $\dmm$).  We have
one or more parametric models for the relationship,
$\ytrue \approx f(\xtrue;\fparams)$, with parameters $\fparams$ (we
sometimes suppress the parameter dependence below to simplify notation).

The data for the analysis are from post-processing outputs from 
deterministic numerical calculations of the binary black hole merger.  
A complex computation produces
initial data (ID) targeting a specified value of $\xtrue$,
but the actual value of $\xtrue$ that the generated ID corresponds to
necessarily differs from the target value.
A processing algorithm estimates the actual value to be $x$.
Evolution of the ID produces high-dimensional
outputs that are processed to produce the computed response, $y$,
that estimates the result, $\ytrue$, that would be obtained by
solving the PDEs exactly.
A set of $(x,y)$ pairs
constitutes the basic data we must use to
find $f(\xtrue;\fparams)$.

A variety of parameters govern the accuracy of the ID, evolution, and
processing algorithms.  These are summarized via a resolution level $\k$
(defined in Sec.~\ref{sec:Methods})
assigned to each $(x,y)$ pair, with the $x$ and $y$ values likely
to be closer to the $\xtrue$ and $\ytrue$ values when $\k$ is larger.
For every choice of ID, we have results for multiple values of $\k$,
comprising repeated measurements of $(\xtrue,\ytrue)$ of varying accuracy.

We have developed 
a Bayesian nonlinear measurement error model for the
analysis.\footnote{
For introductions aimed at physicists,
see Refs.~\cite{J03-PTLOS,S06-DABayesTut,G10-BayesPhySci} for Bayesian
inference and Ref.~\cite{L13-BayesRetroMLMs} for multilevel
Bayesian modeling. Multilevel measurement error models inspiring
our approach here are covered in Refs.~\cite{C+06-MsmtErr,G95-RepeatMsmts}.
}
Letting the index $\n$ label the choice of ID, the model
specification is:
\begin{align}
x_{\n\k} &= \xtrue_\n + \epsilon_{\n\k},\label{xeqn}\\
y_{\n\k} &= \ytrue_\n + \delta_{\n\k}\label{yeqn}\\
   &= f(\xtrue_\n;\fparams) + \Delta_\n + \delta_{\n\k},\label{y-f-Delta}
\end{align}
for $N$ ID cases ($\n=1$ to $N$), and $\k\in L_\n$ for ID case $\n$,
where $L_\n$ denotes a set of levels
run for case $\n$ (for most
cases, $L_\n=\{2,3,4\}$, but for runs targeting $\chi_i=\pm 0.44$,
$L_\n=\{1,2,3\}$).  Here $\epsilon_{\n\k}$ and $\delta_{\n\k}$ 
denote {\em level error} terms reflecting the difference
between numerical results at finite resolution and the actual
solution to the differential equations we are studying.
For Eq.~\eqref{y-f-Delta}\ we set
$\ytrue_\n = f(\xtrue_\n;\fparams) + \Delta_\n$, where $\Delta_\n$ is
a {\em discrepancy}
term representing the difference
between the true response and the prediction based on the fitting
function.

To complete the model we must assign (prior) probability density functions (PDFs),
i.e. \emph{priors},
to a number of random variables:
the level errors, $\epsilon_{\n\k}$ and
$\delta_{\n\k}$; the latent predictor variables, $\xtrue_\n$; and the 
latent discrepancy variables, $\Delta_\n$.

We assign independent, zero-mean normal distributions to the level error 
terms, $\epsilon_{\n\k}$ and $\delta_{\n\k}$, with standard deviations
$\sigma_x/\alpha_\k$ and $\sigma_y/\alpha_\k$ (respectively).  We assign
$\alpha_\k$ scale factors to capture the notion that we expect the
errors to be smaller (on average) for higher levels.  For the
calculations here, we took $\alpha_\k = (1/2)^{4-\k}$, so the
standard deviations for the highest-resolution $\k=4$ results
are $\sigma_x$ and $\sigma_y$, and the error scales double for
each decreasing level.  We did not explore this assignment
except to verify that this choice has a much higher likelihood
than taking $\alpha_\k=1$, i.e., the data themselves show clear
evidence for convergence as $\k$ grows.
Although in principle we could let the error scale be different
for each ID case, for simplicity we assign a common error scale across ID
cases; the modest amount of data we have does not indicate a strong
variation of error scale with ID.
We adopt normal distributions, partly for convenience, but also because we
are modeling relationships between scalar quantities calculated from
high-dimensional computational outputs with complicated algorithms.
Presuming the final errors result from numerous additive contributions whose
uncertainties have finite variance, the central limit theorem motivates the
normal choice.

We assign informative but relatively broad priors
for the $\xtrue_\n$ values,
reflecting the ability to produce ID corresponding to a $\xtrue_\n$ value
close to a desired target value, $\mu_\n$.  The priors are normal with
means $\mu_\n$ (equal to the target value for ID case $\n$) and
common standard deviation $w = 0.002$, reflecting the typical
change in mass and spin as a result of the initial relaxation
(as seen in Fig.~\ref{fig:AhVsTime}).
  These values
do not strongly impact the results.

We also assign independent, zero-mean normal distribution PDFs to the
discrepancy terms, with common standard deviation $\sigma_\Delta$.
The quantity $\sigma_\Delta$ represents the typical scale of
systematic error magnitude in the model.
A more flexible and realistic choice would be to assign a correlated
Gaussian process prior over the space of discrepancy {\em functions},
$\Delta(\xtrue_\n)$, and to identify $\Delta_\n = \Delta(\xtrue_\n)$.
This would resemble the practice in the literature on Bayesian
emulation of input/output response surfaces, the prevailing approach
in the literature on the statistical analysis of the results of
deterministic numerical simulations
(see, e.g., \cite{KO01-BayesEmulation,H+04-BayesEmulation}).  But the
goal of that literature is not to find simple and tractable 
fitting functions; it instead builds nonparametric emulators that,
while simpler than the simulators being emulated, are still
computationally nontrivial.  Moreover, the vast majority of
existing work on emulation addresses cases with precisely known
inputs, which is not the case here; uncertainty in the predictor
significantly complicates implementation of Gaussian process
regression~\cite{MR11-GPInputNoise,DBC11-GPUncertainData}.
The independent normal PDF for
$\Delta_\n$ will enable us to invoke a simple approximation leading to
analytical results.

Finally, we adopt flat priors for the fitting function model
parameters, $\fparams$.

The conditional dependency structure of such a multilevel model
can be represented by a directed acyclic graph (DAG).
A graphical model of this type can be readily coded in a DAG-oriented
statistical modeling language (e.g., WinBUGS or JAGS) to enable
Bayesian computation via Markov chain Monte Carlo (MCMC) posterior sampling.
Here the focused goal (finding a simple fitting function) and the
small uncertainties in the level error terms (well below 1\% for
nonzero spins) motivated an analytical approach based on linearization
of $f(\xtrue)$.  This lets us avoid the complexity
of MCMC, producing a fast algorithm that is relatively simple to use.

\begin{figure}
\centerline{\includegraphics[width=3.4in]{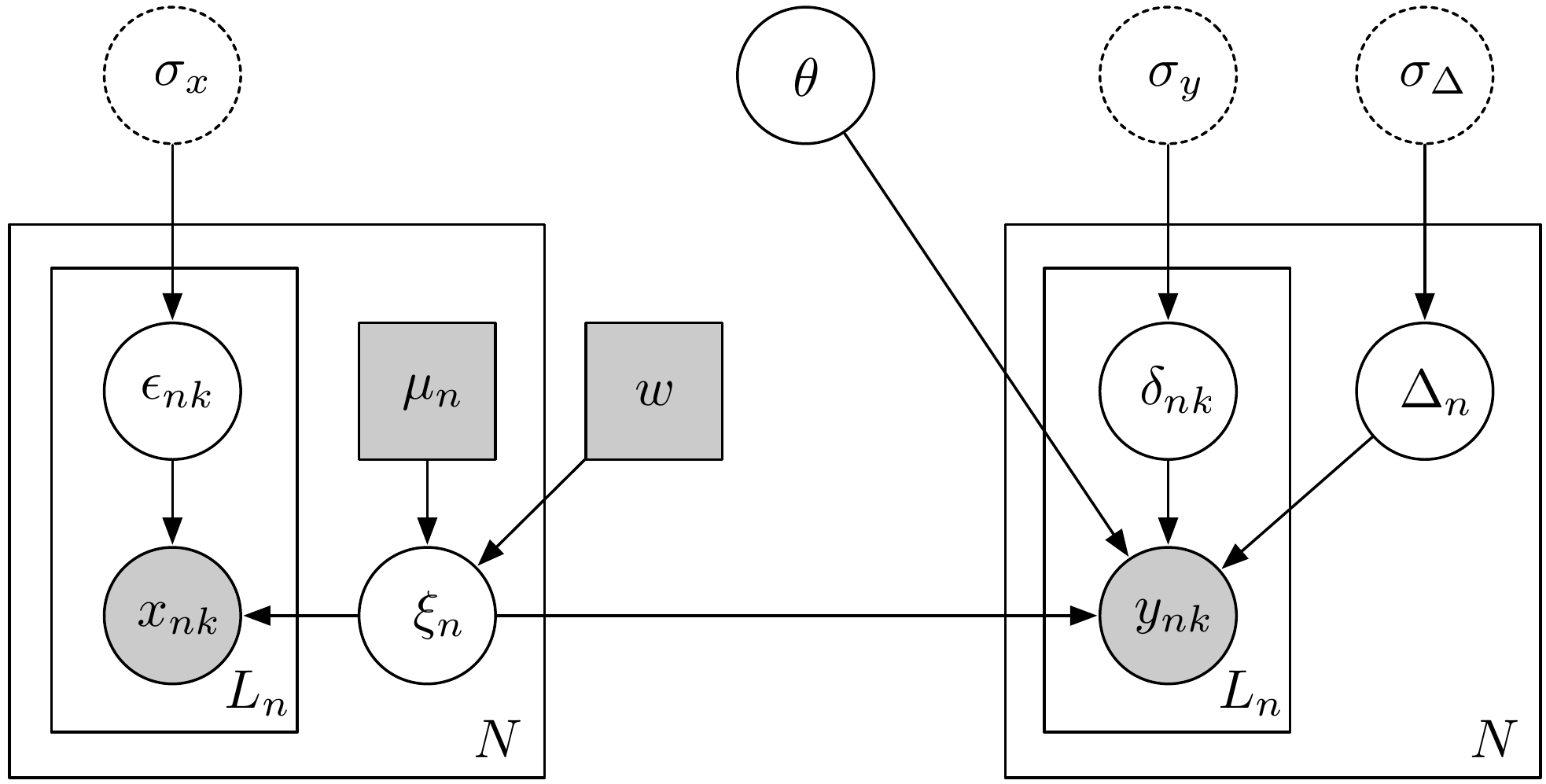}}
\caption{Directed acyclic graph displaying the conditional dependence
structure of the Bayesian nonlinear measurement error model adopted
for the fitting function analysis.  See text for a detailed description.}
\label{fig:dag}
\end{figure}

Figure~\ref{fig:dag} shows the DAG for our model.
Circles denote
random variables (RVs, uncertain quantities with assigned or computed PDFs).
Shaded circles are the data $(x,y)$, and shaded squares are fixed constants
that help define the model.
We marginalize over the error RVs ($\epsilon_{nk}$,
$\delta_{nk}$, and $\Delta_n$) and the uncertain input
variables ($\xi_n$), and then we solve for the remaining
non-shaded variables simultaneously.
The plates (enclosing boxes) denote parts of the graph that
are replicated as indicated by the quantity in the lower right
corner of each plate.


Dashed circles indicate RVs that play
the role of {\em hyperparameters}, i.e., parameters defining
prior PDFs for lower-level RVs.  Formally, we could
account for uncertainty in the hyperparameters by assigning them
priors of their own and marginalizing
over them (the {\em hierarchical Bayes} approach).  As a simpler
approximation, we optimized these hyperparameters (the {\em empirical
Bayes} approach), using constant prior PDFs for them.  

Directed edges (arrows) in a Bayesian network DAG are used to indicate the
dependency structure.
The top-level RVs have no dependencies (incoming arrows); their PDFs would
be specified a priori for a full hierarchical Bayesian analysis.
The conditional PDFs of lower-level RVs depend only on the values
of their dependencies.
The full joint
PDF for all RVs is the product of the prior and conditional PDFs.
Therefore, Fig.~\ref{fig:dag}
indicates that the joint PDF for the RVs comprising our model may be
written
\begin{align}
p(\theta,\sigma_x,&\sigma_y,\sigma_\Delta,\xtrue,\Delta,\epsilon,\delta,x,y) =
    p(\theta) p(\sigma_x) p(\sigma_y) p(\sigma_\Delta) \nonumber\\
 &\times\prod_{\n=1}^N \Bigg[ p(\xtrue_\n|\mu_\n,w)\,p(\Delta_\n|\sigma_\Delta) \vphantom{\prod_{k\in L_\n}}\Bigg.\nonumber\\
 &\times \prod_{k\in L_\n} p(\epsilon_{\n\k}|\sigma_x)
        p(x_{\n\k}|\xtrue_\n,\epsilon_{\n\k}) \nonumber\\
 &\times \Bigg.p(\delta_{\n\k}|\sigma_y) p(y_{\n\k}|\theta,\xtrue_\n,\Delta_\n,\delta_{\n\k}) \Bigg],
\label{full-joint}
\end{align}
where $(\xtrue,\Delta,\epsilon,\delta,x,y)$ is shorthand notation for the
indexed collections of those variables.  Since we are adopting
a constant prior PDF for $\fparams$, and an empirical Bayes treatment
of the hyperparameters $\hypers = (\sigma_x,\sigma_y,\sigma_\Delta)$, the
quantity of interest is the conditional PDF for the data, $(x,y)$, and the
latent parameters, $(\xi,\Delta,\epsilon,\delta)$, given the fitting
function parameters and the hyperparameters,
\begin{align}
p(\xtrue,\Delta,&\epsilon,\delta,x,y|\theta,\hypers)
 = \prod_{\n=1}^N \Bigg[ p(\xtrue_\n|\mu_\n,w)\,p(\Delta_\n|\sigma_\Delta) \vphantom{\prod_{k\in L_\n}}\Bigg.\nonumber\\
 & \times \prod_{k\in L_\n} p(\epsilon_{\n\k}|\sigma_x)
        p(x_{\n\k}|\xtrue_\n,\epsilon_{\n\k}) \nonumber\\
 & \times \Bigg.p(\delta_{\n\k}|\sigma_y) p(y_{\n\k}|\theta,\xtrue_\n,\Delta_\n,\delta_{\n\k}) \Bigg],
\label{cond-data-latents}
\end{align}

The model Eqs.~\eqref{xeqn} and~\eqref{y-f-Delta} imply that the conditional 
PDFs for $x_{\n\k}$ and $y_{\n\k}$ in these equations are $\delta$-functions.
This lets us trivially marginalize over $\epsilon$
and $\delta$, giving a marginal PDF for the remaining variables,
\begin{align}
p(\xtrue,\Delta,& x,y|\theta,\hypers)
 = \prod_{\n=1}^N \Bigg[ p(\xtrue_\n|\mu_\n,w)\,p(\Delta_\n|\sigma_\Delta) \vphantom{\prod_{k\in L_\n}}\Bigg.\nonumber\\
 & \times \prod_{k\in L_\n} p(\epsilon_{\n\k}=x_{\n\k}-\xi_\n|\sigma_x) \nonumber\\
 & \times \Bigg.p(\delta_{\n\k}=y_{\n\k}-f(\xi_\n;\params)-\Delta_\n|\sigma_y)  \Bigg].
\label{level-err-marg}
\end{align}
Marginalizing over $\xtrue$ and $\Delta$ gives the {\em marginal
likelihood function} (the probability for the data, conditioned on
parameter values) for the fitting parameters and hyperparameters,
\begin{align}
&\like_M(\fparams,\hypers) 
 = \prod_{\n=1}^N \int d\xtrue_\n \int d\Delta_\n\,
    \left[ \normal(\xtrue_\n|\mu_\n,w)\,\normal(\Delta_\n|\sigma_\Delta) \vphantom{\prod_{k\in L_\n}}\right.\nonumber\\
 & \left.\times \prod_{k\in L_\n} \normal(x_{\n\k}-\xi_\n|0,\sigma_x)
        \normal(y_{\n\k}-f(\xi_\n;\params)-\Delta_\n|0,\sigma_y)  \right],
\label{L-marg}
\end{align}
where $\normal(z|\mu,\sigma)$ denotes the normal distribution PDF for $z$ with
mean $\mu$ and standard deviation $\sigma$,
\begin{equation}
\normal(z|\mu,\sigma) = \frac{1}{\sigma\sqrt{2\pi}} e^{-(z-\mu)^2/2\sigma^2}.
\label{norm-def}
\end{equation}
When $f(\xtrue)$ is a nonlinear function of $\xtrue$, the $\xtrue$ integral
in Eq.~\eqref{L-marg} is in general intractable.  However, the $x$ and $y$ errors are
small, so we expect a local linear approximation of $f(\xtrue)$ to be
very accurate over regions of $\xi_\n$ that have significant probability
density.  So we use
\begin{equation}
f(\xtrue_\n;\fparams) \approx f(\tilde\xtrue_\n;\fparams)
  + (\xtrue_\n - \tilde\xtrue_\n) f'(\tilde\xtrue_\n;\fparams)
\label{f-lin}
\end{equation}
in Eq.~\eqref{L-marg}, where $f'(\xtrue;\fparams)$ denotes the derivative of the
fitting function with respect to $\xtrue$, and $\tilde\xtrue_\n$ is a fixed
reference value of $\xtrue_\n$ based on the $x_{\n\k}$ values for a particular
$n$ (we use a weighted mean of the $x_{\n\k}$).
With this linearization, the
integrals in the marginal likelihood function can be performed analytically.

We estimate the parameters for a candidate fitting function by maximizing
the marginal likelihood function over both $\fparams$ and $\hypers$:
\begin{equation}
(\hat\fparams, \hat\hypers) = \arg\max \like_M(\fparams,\hypers).
\end{equation}
For the fitting functions studied here, the $\fparams$ dependence of
the marginal likelihood function is approximately multivariate Gaussian.
To quantify the $\fparams$ uncertainties, we calculate the observed
Fisher information matrix (with $\hypers$ fixed at $\hat\hypers$),
\begin{equation}
I_{\alpha\beta} 
  = \frac{\partial^2}{\partial\fparams_{\alpha}\partial\fparams_{\beta}}
    \like_M(\fparams,\hypers) \Big|_{\hat\fparams, \hat\hypers},
\label{info-matrix}
\end{equation}
where $\fparams_\alpha$ denotes the $\alpha$th parameter of the fitting
function.  The posterior PDF for $\fparams$ (conditional on $\hat\hypers$)
is then approximately a multivariate normal PDF with mean $\hat\fparams$ and
covariance matrix $\Sigma = I^{-1}$.

To predict the value of the response at a specified value of $\xtrue$,
we calculate an approximate predictive distribution (also conditioned
on $\hat\hypers$) using the multivariate normal PDF and the delta
method (propagation of errors).  The model assumes the response
is given by the sum of the fitting function and a discrepancy
term with zero mean.  The most probable value of the
response is simply $f(\xtrue;\hat\fparams)$.  There are two components
to the uncertainty in the prediction.  One comes from propagating the
$\fparams$ uncertainty (accounting for correlations between the
parameters, which can be large).  The resulting standard deviation
in the fitting function evaluated at $\xtrue$ is $\sigma_f(\xtrue)$, satisfying
\begin{equation}
\sigma^2_f(\xtrue)
  = \sum_{\alpha\beta} \frac{\partial f(\xtrue;\fparams)}{\partial\fparams_\alpha}
    \Sigma_{\alpha\beta} \frac{\partial f(\xtrue;\fparams)}{\partial\fparams_\beta}.
\label{sigma-f}
\end{equation}
The full uncertainty in the predicted response must also account
for the uncertainty in the discrepancy term, which is given by the
hyperparameter $\sigma_{\Delta}$ that we estimate from the data.  The
full uncertainty in the prediction is
\begin{equation}
\sigma_{\rm tot}(\xtrue) = \sqrt{\sigma^2_f(\xtrue) + \hat\sigma_{\Delta}^2}.
\label{sigma-tot}
\end{equation}
This calculation ignores the uncertainty in the value of $\sigma_{\Delta}$,
but that uncertainty is relatively small in our calculations.

To compare rival parametric fitting functions, a formal model comparison
could be implemented, e.g., using Bayes factors (which would require
assigning normalized priors to $\fparams$ for each candidate fitting
function, and integrating the product of the prior and the marginal
likelihood function over $\fparams$), or an information criterion
such as the Bayesian information criterion (BIC) or the Akaike information
criterion (AIC).  The BIC and AIC rank models according to their
maximum likelihoods, penalized by a term depending on the number
of parameters in each model (and the sample size in the case of
the BIC).  These criteria were developed for comparing simple
parametric models, not multilevel models with many latent parameters.
We adopt a less formal approach here.  We simply calculate the
logarithm of ratios of the maximum marginal likelihood function.
For the models we consider, the log-ratio for the best model vs.\ the
next-best competitor is large (well over 10), far larger than the
typical penalty terms in information criteria, so the choice of
best model is unambiguous.

\begin{acknowledgments}
We thank Matt Giesler, Tony Chu, and Bryant Garcia for providing data from
their simulations.
We gratefully
acknowledge support from the Sherman Fairchild Foundation; from NSF
grants PHY-0969111 and PHY-1005426 at Cornell; and from NSF grants
PHY-1068881, PHY-1005655, and DMS-1065438  at Caltech.  
Loredo's work for this project was supported
by NSF grant AST-0908439 and NASA grant NNX09AK60G.
Simulations used in this work
were computed with SpEC~\cite{SpECwebsite}.  Computations
were performed on
SHC at Caltech, which is supported by the Sherman
Fairchild Foundation; on
the Zwicky cluster at Caltech, which is supported by
the Sherman Fairchild Foundation and by NSF award PHY-0960291; on the
NSF XSEDE network under grant TG-PHY990007N; and on the GPC
supercomputer at the SciNet HPC Consortium~\cite{scinet}. SciNet is
funded by: the Canada Foundation for Innovation under the auspices of
Compute Canada; the Government of Ontario; Ontario Research
Fund--Research Excellence; and the University of Toronto.
\end{acknowledgments}

\bibliography{References/References}
\end{document}